\documentclass{aa}  
\usepackage{graphicx}
\usepackage{txfonts}
\usepackage{lscape}

\begin{document}

\title{Infrared to X-ray observations of PKS 2155-304 in a low state}

\author{L. Foschini\inst{1}, A. Treves\inst{2}, F. Tavecchio\inst{3}, D. Impiombato\inst{4}, G. Ghisellini\inst{3}, S. Covino\inst{3}, G. Tosti\inst{4}, M. Gliozzi\inst{5}, V. Bianchin\inst{1}, G. Di Cocco\inst{1}, G. Malaguti\inst{1}, L. Maraschi\inst{3}, E. Pian\inst{6}, C. M. Raiteri\inst{7}, R. M. Sambruna\inst{8}, G. Tagliaferri\inst{3} \and M. Villata\inst{7}
}

%\offprints{L. Foschini}

\institute{INAF/IASF-Bologna, Via Gobetti 101 - 40129 Bologna (Italy)\\\email{foschini@iasfbo.inaf.it}
\and
Dipartimento di Scienze, Universit\`a degli Studi dell'Insubria, Via Valleggio 11, 22100, Como (Italy)
\and
INAF, Osservatorio Astronomico di Brera, Via Bianchi 46, 23807, Merate (Italy)
\and
Osservatorio Astronomico, Universit\`a di Perugia, Via B. Bonfigli, 06126 Perugia (Italy)
\and
George Mason University, 4400 University Drive, Fairfax, VA, 22030, USA
\and
INAF, Osservatorio Astronomico di Trieste, Via G.B. Tiepolo 11, 34131, Trieste (Italy) 
\and
INAF, Osservatorio Astronomico di Torino, Via Osservatorio 20, 10025, Pino Torinese (Italy)
\and
NASA Goddard Space Flight Center, Code 661, Greenbelt, MD 20771, USA
}

\date{Received April 15, 2008; accepted April 29, 2008}

  \abstract
  {} 
  %leave it empty if necessary  
   {Our goal is to understand the nature of blazars and the mechanisms for the generation of high-energy $\gamma-$rays, through the investigation of the prototypical blazar PKS~$2155-304$, which shows complex behaviour.}
   {We analyze simultaneous infrared-to-X-ray observations obtained with \emph{XMM-Newton} and \emph{REM} on November $7$, $2006$, when the source was in a low X-ray state. We perform a comparative analysis of these results with those
obtained from previous observations in different brightness states.}
   {We found that the peak of the synchrotron emission moved from ultraviolet to optical wavelengths and the X-ray spectrum is best fit with a broken power law model with $\Gamma_2 \sim 2.4$ harder than $\Gamma_1 \sim 2.6$ and a break at about $3.5$~keV. This suggests that the soft X-rays ($E < 3.5$~keV) are related to the high-energy tail of the synchrotron emission, while the hard X-rays ($E > 3.5$~keV) are from the energy region between the synchrotron and inverse-Compton humps. The different variability at energies below and above the break strengthens this hypothesis. Our results also stress the importance of monitoring this source at both low and high energies to better characterize its variability behaviour.}
  {}
  

\authorrunning{L. Foschini et al.}

\maketitle

\section{Introduction}
Blazars are active galactic nuclei having an energy spectrum dominated by the relativistically beamed radiation coming from a jet pointed at the Earth and moving at speeds close to $c$ (Urry \& Padovani 1995). The long-term behaviour of these sources is characterized by periods with low high-energy flux and periods with emission particularly strong and violent at X- and $\gamma-$rays. The spectral energy distribution (SED) generally changes both in shape and normalization during outbursts, in a complex way.

PKS~$2155-304$ ($z=0.116$) is one of these enigmatic sources. Although it has been observed since the seventies (earliest observations are from \emph{Ariel V}, Cooke et al. 1978) with many space and ground observatories at all wavelengths, it still shows unexpected behaviour, like the giant TeV flares, which occurred at the end of July $2006$ (Aharonian et al. 2007, Foschini et al. 2007, Mazin \& Lindfors 2007, Sakamoto et al. 2008). The largest outburst, recorded on July $28$, was characterized by an average flux -- in the $E > 200$~GeV energy band -- 7 times the Crab flux and several peaks that doubled the mean value on timescales of the order of a few minutes (Aharonian et al. 2007). This giant flare was followed 2 days later by another strong flare, which was covered by \emph{Swift} observations (Foschini et al. 2007). The \emph{Swift} follow-up lasted about one month and measured the decaying of the X-ray flux.
The main anomalies of these violent outbursts were the very short timescale of TeV flux variations (Begelman et al. 2008, Ghisellini \& Tavecchio 2008) and the fact that, despite these strong variations at TeV energies, little or no spectral changes were observed at any wavelengths (Foschini et al. 2007). 

\begin{figure}[!t]
\centering
\includegraphics[angle=270,scale=0.34,clip,trim=37 0 0 0]{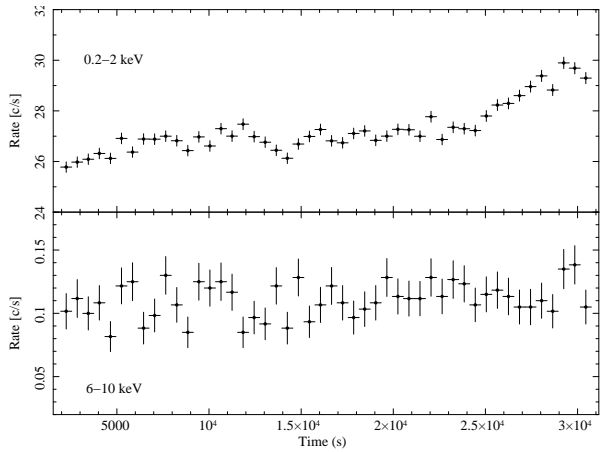}
\caption{EPIC-pn lightcurves in the $0.2-2$~keV band \emph{(top panel)} and $6-10$~keV band \emph{(bottom panel)} with $600$~s time bins; time starts on November $7$, $2006$, $00:33:07$~UT.}
%\caption{\emph{(left panel)} EPIC-pn lightcurve in the $0.5-10$~keV band with $600$~s time bins; time starts on November 7, 2006, 00:33:07 UT. \emph{(right panel)} EPIC-pn spectrum fit with the broken power-law model with residuals in terms of sigmas with error bars of size one (bottom panel).}
\label{fig:epic}
\end{figure}

The \emph{REM} telescope observed the source from the end of August to the end of $2006$. During the local optical minimum, on November $7$, $2006$, \emph{XMM-Newton} observed the source for about $30$~ks (see Fig.~\ref{fig:epic}). Here we report the analysis of these data and a comparison with previous X-ray and optical observations carried out when PKS~$2155-304$ was in a high activity state. 

\section{Data analysis}

\subsection{XMM-Newton}
\emph{XMM-Newton} observed PKS~$2155-304$ on November $07$, $2006$ (ObsID $0411780101$) as a routine calibration observation. Data of EPIC-pn (Str\"uder et al. 2001), set in small window mode, and OM (Mason et al. 2001) have been processed, screened and analyzed by using the same procedures described in Foschini et al. (2006a), but with \texttt{SAS v 7.1.0} and the calibration files as of July $16$, $2007$. Data of EPIC-MOS (Turner et al. 2001) were not analyzed because they were affected by pile-up. The EPIC-pn alone guarantees very high statistics and it is not affected by pile-up at these flux levels.

\begin{figure*}[!t]
\centering
\includegraphics[scale=0.42,clip,trim=0 50 0 0]{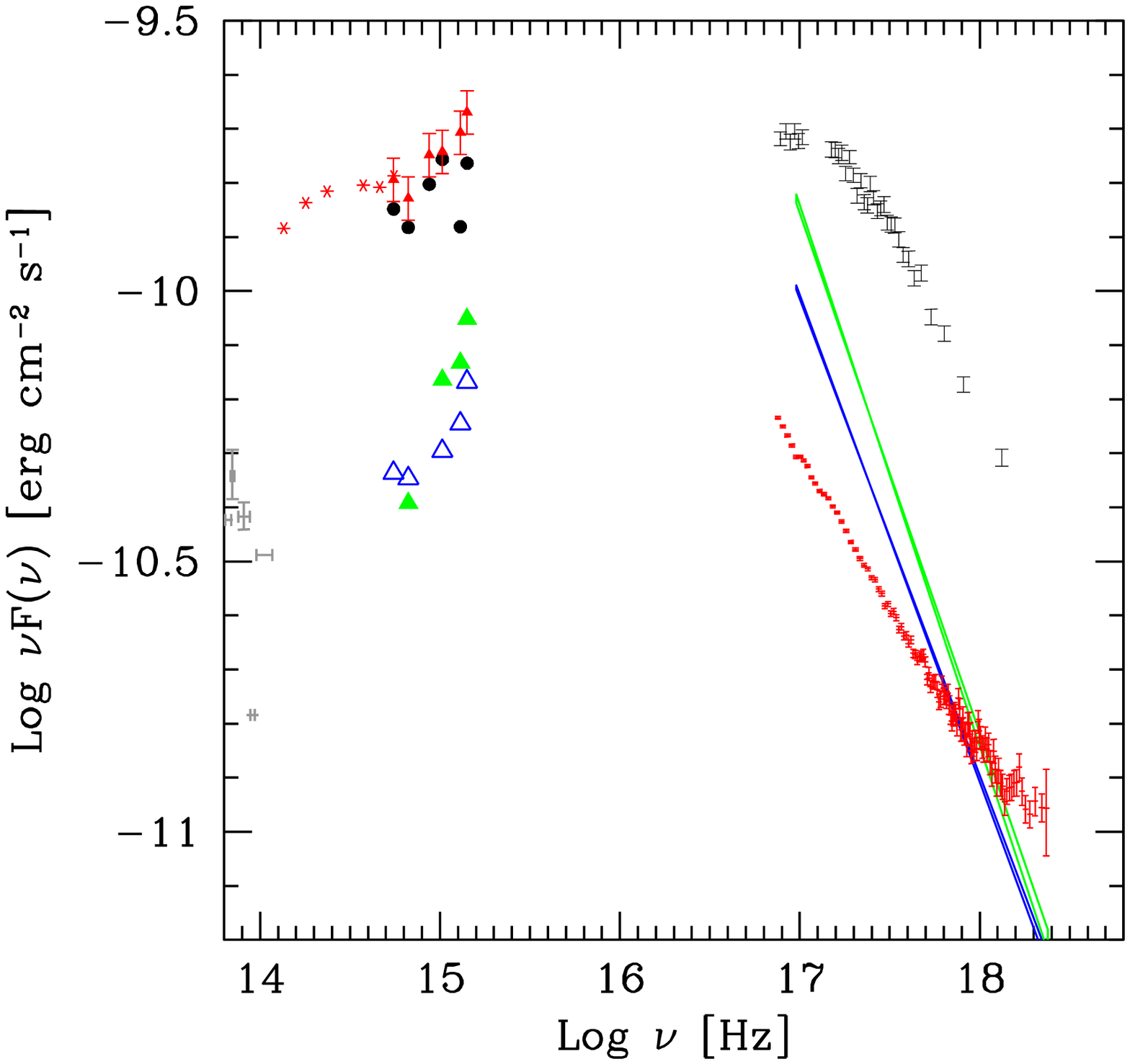}
\includegraphics[scale=0.42,clip,trim=0 50 0 0]{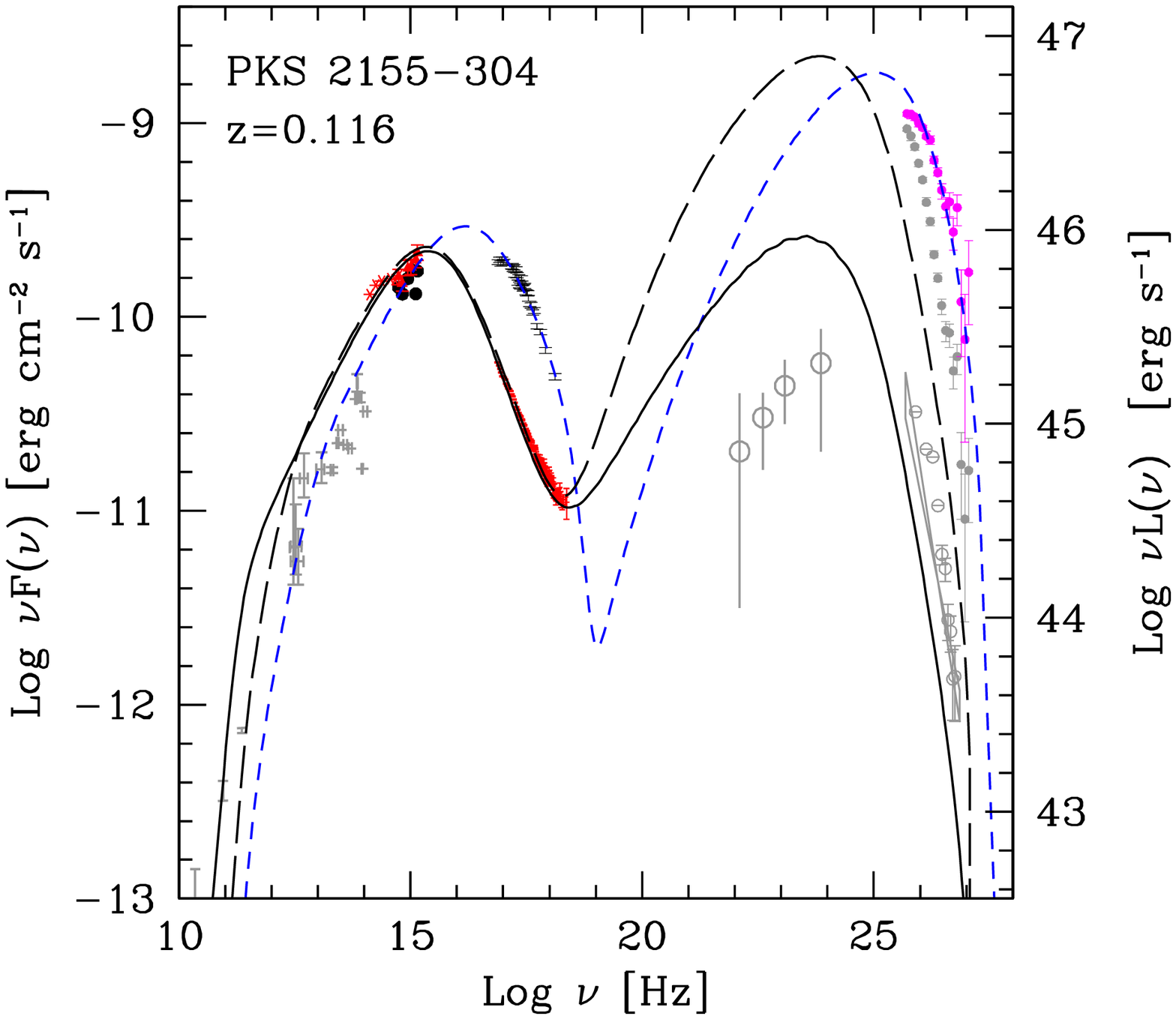}
\caption{\emph{(left panel)} Zoom of the SED of PKS~$2155-304$ built with simultaneous near-infrared/optical data from \emph{REM} (red asterisks) and optical-to-X-ray data from \emph{XMM-Newton} (red triangles and bars). Other colors refer to \emph{XMM-Newton} data from Foschini et al. (2006a). \emph{(right panel)} Complete SED with observed TeV data (grey filled circles) from Aharonian et al. (2007), while magenta filled circles representing TeV data corrected for absorption by using the low SFR model by Kneiske et al. (2004). The synchrotron self-Compton model used to fit the data of November $7$, $2006$ is represented with a black continuous line (low SSC) and a black dashed line (high SSC), while the magenta short-dashed line is the fit to the July $30$, $2006$ data from Foschini et al. (2007).}
\label{fig:sed}
\end{figure*}

The EPIC-pn spectrum (not shown, but see Fig.~\ref{fig:sed}) is well fitted ($\tilde{\chi}^2=1.04$ for $1010$ degrees of freedom) with a broken power-law model with fixed Galactic absorption ($N_{\rm H}=1.48\times 10^{20}$~cm$^{-2}$, Kalberla et al. 2005). The parameters of the broken power-law model are: $\Gamma_1=2.563\pm 0.004$, $\Gamma_2=2.39\pm 0.06$ and $E_{\rm break}=3.5_{-0.6}^{+0.5}$~keV. The flux is $2.3\times 10^{-11}$~erg~cm$^{-2}$~s$^{-1}$ in the $2-10$~keV energy band. The broken power-law model is statistically required ($Ftest > 99.99$\%) with respect to the single power-law or the log-parabola models (cf, e.g., Massaro et al. 2008).

The EPIC-pn lightcurve in the $0.2-2$~keV energy band (Fig.~\ref{fig:epic}, \emph{top panel}) shows significant variability ($\chi^2$ probability of constancy $\rightarrow 0$) with RMS $3.5\pm 0.4$\%, while the $6-10$~keV band lightcurve (Fig.~\ref{fig:epic}, \emph{bottom panel}) is consistent with a constant ($\chi^2$ probability of constancy $0.45$) with an upper limit ($3\sigma$) for the RMS equal to $11$\%.

The OM observed magnitudes (error $\pm 0.1$~mag) are: $V_{543 \rm nm} = 12.7$, $B_{450 \rm nm} = 13.1$, $U_{344 \rm nm} = 12.2$, $UVW1_{291 \rm nm} = 11.9$, $UVM2_{231 \rm nm} = 11.8$, $UVW2_{212 \rm nm} = 11.9$. 

Comparing these values with those from previous \emph{XMM-Newton} observations reported in Foschini et al. (2006a), we observe a lower X-ray flux (even though not the lowest), an inverted broken power law ($\Gamma_2 < \Gamma_1$), and a brighter optical/UV state. A comparison with the observations carried out by \emph{Swift} during the giant TeV flare on July $2006$ (Foschini et al. 2007) shows a similar optical state, but a lower X-ray flux. The different variability in the soft and hard energy bands favours the hypothesis of a different origin of the two components, as generally occurs in low-frequency BL Lac Objects (LBL or intermediate blazars), like, e.g., S5 0716+71 (Foschini et al. 2006b) or ON 231 (Tagliaferri et al. 2000).

\begin{figure}[!t]
\centering
\includegraphics[scale=0.55,clip,trim=0 0 0 17]{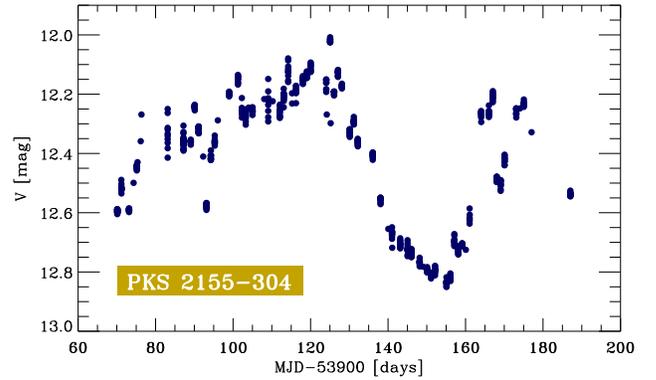}
\caption{Lightcurve (V filter) obtained from \emph{REM} observations during the period August-December $2006$. The average error is $0.04$ mag. \emph{XMM-Newton} observation starts at MJD$=54046.023$ (Nov. $7$, $2006$, $00:33:07$ UT). See also Impiombato et al. (2008).}
\label{fig:rem}
\end{figure}

\subsection{REM}
The \emph{Rapid Eye Mount (REM)} is an automatic telescope equipped with two cameras operating in the near-infrared ($z^\prime$, $J$, $H$, $K$) and optical ($I$, $R$, $V$) frequencies (Chincarini et al. 2003, Covino et al. 2004). 

The observations of PKS~$2155-304$ started on August $23$, $2006$ and ended on December $19$, $2006$ (Fig.~\ref{fig:rem}). Images obtained from \emph{REM} have been corrected for bias, dark and flat-field and then analyzed using \texttt{GAIA} software\footnote{http://docs.jach.hawaii.edu/star/sun214.htx/sun214.html}. Stars from the Landessternwarte Heidelberg-K\"onigstuhl (optical)\footnote{http://www.lsw.uni-heidelberg.de/projects/extragalactic/charts/} and 2MASS catalogs (near-infrared)\footnote{http://www.ipac.caltech.edu/2mass/} have been used for calibration. More details on the \emph{REM} data analysis can be found in Impiombato et al. (2008).

The average magnitudes observed during November $7$, $2006$, between $00:53$ and $02:00$ UT are: $V_{0.54 \mu\rm m}=12.73\pm 0.04$, $R_{0.65 \mu\rm m}=12.40\pm 0.02$, $I_{0.80 \mu\rm m}=11.96\pm 0.03$, $J_{1.27 \mu\rm m}=10.98\pm 0.03$, $H_{1.67 \mu\rm m}=10.31\pm 0.03$, $K_{2.22 \mu\rm m}=9.60\pm 0.02$. The magnitudes (also in the case of \emph{XMM-Newton}/OM data) have been dereddened using $A_V=0.071$ and the extinction laws by Cardelli et al. (1989). Then, they were converted into fluxes with standard formulae.

The lightcurve (Fig.~\ref{fig:rem}) exhibits a maximum of $V=12$ mag on MJD$=54025$, brighter than the optical state corresponding to the July TeV flares ($V=12.6$ on July $30$, Foschini et al. 2007). Our \emph{REM} monitoring during $2005$ (Dolcini et al. 2007a,b) never detected the source in such a bright state both in optical and near-infrared. The \emph{XMM-Newton} observations we are considering here are close to the optical minimum in the August-December $2006$ period, which in turn is close to the state of July $30$. This underscores the complexity of the broadband behaviour.

\section{Spectral Energy Distribution}
The spectral energy distribution (SED) of PKS~$2155-304$, assembled with the data described above, together with previous observations, is shown in Fig.~\ref{fig:sed}. In the present work, to fit the SED we used a synchrotron self-Compton (SSC) model slightly different from the one adopted in Foschini et al. (2007). In the SSC model used here, the emitting region is a sphere with radius $R$ moving with bulk Lorentz factor $\Gamma$, with a tangled and uniform magnetic field $B$. The viewing angle of the observer is $\theta$, which in turn means a Doppler factor $\delta$. The purely phenomenological distribution of the emitting relativistic electrons is described by a broken power-law model with normalization $K$ and indices $n_1$ from $\gamma_{\rm min}$ to $\gamma_{\rm break}$ and $n_2$ above the break up to $\gamma_{\rm max}$. A full description of the model can be found in Maraschi \& Tavecchio (2003). The main difference is the treatment of the Klein-Nishina cross section: in the previous work, we adopted a truncated expression, while, in the present case, we used the full treatment. For completeness and to allow a direct comparison with the present results, we also fit with this SSC model the data of the TeV flare that occurred on July $30$, $2006$ (short dashed line in Fig.~\ref{fig:sed}). We use the time-averaged TeV spectrum measured by the \emph{H.E.S.S.} telescope (Aharonian et al. 2007), corrected for the extragalactic absorption by using the ``Low SFR'' model of Kneiske et al. (2004). Although low, these values for the level of the extragalactic background light (EBL) are in agreement with the observations (e.g. Aharonian et al. 2006, Mazin \& Raue 2007). Compared to Foschini et al. (2007), the present model shows a slight shift of the inverse-Compton peak, an increase of the Doppler factor $\delta$ and a decrease of the magnetic field $B$, but the main conclusions are not altered by this change. The parameters derived from the updated model of July $30$ SED are summarized in Table~\ref{param}.

Using the new version of the SSC model described above during the outburst on July $30$, $2006$, the Doppler factor required by the fit is $\delta=50$, in agreement with the values derived by Begelman et al. (2008) and Ghisellini \& Tavecchio (2008). On the other hand, this value is rather lower than that derived by Finke et al. (2008), who proposed a Doppler factor as large as $100-200$. The main reason for this difference is the higher level of the EBL used by Finke et al. (2008), who fixes the peak of the SSC component around $\nu_C = 3\times 10^{25}$ Hz, and, as discussed
in Tavecchio \& Ghisellini (2008), for a fixed synchrotron component, the value of $\delta$ scales as $\nu_C^{1/2}$.

The SED built with the optical and the X-ray data collected on November $7$, $2006$, presents a very interesting feature: the position of the synchrotron peak is robustly constrained around $10^{15}$ Hz, below the values usually observed. However, the lack of information on $\gamma-$rays on the same days prevents us from fully constraining the model parameters. Therefore, we report two possible models, characterized by similar synchrotron emission, but different levels of the SSC component (cf Fig.~\ref{fig:sed}: high, long dashed line; low, continuous line). The difference between the two models is mainly in the value of the magnetic field ($B=0.1$~G and $B=0.28$~G, for high SSC and low SSC, respectively) and the value of $\gamma_{\rm min}$ ($600$ and $50$, high SSC and low SSC, respectively). The main effect of a large $\gamma_{\rm min}$ is to produce a ``narrow'' SSC component. This is required for the high SSC case in order to avoid an over-production of the observed X-rays in the medium-hard band.

As shown in Table~\ref{param}, all models require rather similar values of the power carried by the jet (calculated assuming one proton per relativistic electron). This conclusion is not unique: the power for the model of July 30 is not strongly constrained, greater power could be derived by assuming a lower $\gamma_{\rm min}$ without violating the observed data. 

\begin{table}[!t]
\begin{center}
\caption{Parameters for the SSC model. See the text for more details.} 
\begin{tabular}{lccc}
\hline
Parameter                              & July 30    & November 7 & November 7\\
{}                                     & {}         & (low SSC)  & (high SSC)\\
\hline
$R$ ($10^{15}$~cm)                     & $5$        & $17.5$               &$17.5$\\
$K$ ($10^{5}$~cm$^{-3}$)               & $1.0$      & $0.165$              &$0.85$\\
$\gamma_{\rm min}$   ($10^3$)          & $1$        & $0.05$               &$0.6$ \\
$\gamma_{\rm break}$ ($10^4$)          & $5$        & $1.5$                &$2.4$\\
$\gamma_{\rm max}$   ($10^5$)          & $5$        & $10$                 &$10$\\
$n_1$                                  & $2$        & $2$                  &$2$\\
$n_2$                                  & $4.12$     & $4.47$               &$4.38$\\
$B$ (gauss)                            & $0.035$    & $0.28$               &$0.1$\\
$\delta$                               & $50$       & $15$                 &$15$\\
$L_{\rm jet}$ ($10^{45}$~erg~s$^{-1}$) & $5.2$      & $3.8$                &$3.9$\\
\hline
\end{tabular}
\label{param}
\end{center}
\end{table}

\section{Discussion}
Being one of the most prominent blazars and among the first to be discovered, PKS~$2155-304$ has been observed numerous times by many satellites and ground-based instruments over the last $\approx 30$ years. These observations have consistently shown that the frequency of the synchrotron peak is generally located at $\approx 10^{16-17}$~Hz and did not change very much over the years (see, for example, Urry et al. 1997, Chiappetti et al. 1999, Foschini et al. 2006a, 2007). Even during the strong TeV flare of July $2006$, the synchrotron peak frequency did not increase to higher values, as generally occurs in high-frequency peaked BL Lacs (HBL). There were flux variations, but negligible spectral changes. On the contrary, Foschini et al. (2007) found that during the second TeV flare the synchrotron peak occurred at $\approx 10^{16}$~Hz, similar to the values obtained during weaker TeV activity.

Interestingly, the observations performed on November $7$, $2006$ indicate a synchrotron peak shift to lower frequencies ($\approx 10^{15}$~Hz), while a hard tail appears in the X-ray spectrum above $\approx 3.5$~keV. Previous claims of a hardening in the X-ray spectrum, generally above $10$~keV, of PKS~$2155-304$ date back to \emph{HEAO1} (Urry \& Mushotzky 1982) and \emph{BeppoSAX} (Giommi et al. 1998). In the former there was only a marginal detection, while for the latter there was no confirmation in subsequent reanalyses (Donato et al. 2005, Giommi et al. 2002). Moreover, the lack of simultaneous optical observations did not provide constraints on the frequency of the synchrotron maximum.

The present high-quality and simultaneous near-infrared-to-X-ray observations allow us to catch PKS~$2155-304$ in a low state, characterized by a shift of the synchrotron peak to low frequencies. According to the best fit of the SSC model (Table~\ref{param}) this corresponds to a deceleration and expansion of the jet ($\delta = 15$ vs $\delta = 50$ during the July outburst; $R=17.5\times 10^{15}$~cm vs $R=5\times 10^{15}$~cm). In addition, it is reasonable to conclude that PKS~$2155-304$ is usually in a high activity phase and only seldom moves to low activity periods, as in the present episode. This means that the jet is emitting almost continuously.

However, a change in the classification of PKS~$2155-304$ is possibly required. Indeed, the spectral characteristics observed on November 2006 resemble those of LBL, like e.g. S5 0716+71 (Foschini et al. 2006b) or ON 231 (Tagliaferri et al. 2000). On the other hand, during the July 2006 flare, it exhibited an inverse-Compton dominance in the SED, which is not typical of HBL. This suggests that blazars could change their spectral type in certain circumstances. It is not clear if this is true for PKS~$2155-304$ only or if it is a general characteristic of blazars. This question can be answered by regularly monitoring and studying specific blazars, which can be considered as test cases. The availability of data in the $\gamma-$ray energy band is crucial, as shown in the present work, to disentangle different models. The forthcoming launch of \emph{GLAST} should cover this lack.

\emph{Note:} A few days after having submitted our manuscript, a work by Zhang (2008), who independently reached conclusions similar to ours, was posted on \texttt{arXiv.org}.

\begin{acknowledgements}
This research has made use of data obtained from the High Energy Astrophysics Science Archive Research Center (HEASARC), provided by NASA's Goddard Space Flight Center and of observations made with the REM Telescope, INAF Chile. LF acknowledges partial support from ASI/INAF Contract I/088/06/0.
\end{acknowledgements}


\begin{thebibliography}{}

\bibitem[2007]{hess2} Aharonian F. et al., 2007, ApJ 664, L71

\bibitem[2006]{hess2} Aharonian F. et al. 2006, Nature, 440, 1018

\bibitem[2008]{BEGELMAN} Begelman M.C., Fabian A.C. \& Rees M.J., 2008, MNRAS 384, L19

\bibitem[1989]{CARDELLI} Cardelli J.A., Clayton G.C., Mathis J.S., 1989, ApJ 345, 245

\bibitem[1999]{chiappetti} Chiappetti L., Maraschi L., Tavecchio F., et al., 1999, ApJ 521, 552

\bibitem[2003]{REM} Chincarini G. et al., 2003, The Messenger 113, 40

\bibitem[1978]{ARIELV} Cooke B.A. et al., 1978, MNRAS 182, 489

\bibitem[2004]{REM2} Covino S., Zerbi F.M., Chincarini G., et al., 2004, AN 325, 543

\bibitem[2007a]{DOLCINI1} Dolcini A., Farfanelli F., Ciprini S., et al., 2007a, A\&A 469, 503

\bibitem[2007b]{DOLCINI1} Dolcini A., Farfanelli F., Ciprini S., et al., 2007b, A\&A 476, 1219

\bibitem[2005]{DONATO} Donato D., Sambruna R.M. \& Gliozzi M., 2005, A\&A 433, 1163

\bibitem[2008]{dermer} Finke J. et al. 2008, ApJ, submitted [\texttt{arXiv:0802.1529}]

\bibitem[2006a]{foschini1} Foschini L., Ghisellini G., Raiteri C.M., et al., 2006a, A\&A 453, 829

\bibitem[2006b]{foschini3} Foschini L., Tagliaferri G. Pian E., et al., 2006b, A\&A 455, 871

\bibitem[2007]{foschini2} Foschini L., Ghisellini G., Tavecchio F., et al., 2007, ApJ 657, L81

\bibitem[2008]{GT} Ghisellini G. \& Tavecchio F., 2008, MNRAS 386, L28

\bibitem[1998]{GIOMMI1} Giommi P., Fiore F., Guainazzi M., et al., 1998, A\&A 333, L5

\bibitem[2002]{GIOMMI2} Giommi P. et al., 2002, In: ``Blazar Astrophysics with BeppoSAX and Other Observatories'', Eds. P. Giommi, E. Massaro, G.G.C. Palumbo. ESA-ESRIN, p. 63 [\texttt{arXiv:astro-ph/0209596}]

\bibitem[2005]{LAB} Kalberla P.M.W. et al., 2005, A\&A 440, 775

\bibitem[2004]{TEVABSORPTION} Kneiske T.M. et al., 2004, A\&A 413, 807

\bibitem[2008]{REM4} Impiombato D., Tosti G., Treves A., et al., 2008, In: Blazar variability across the electromagnetic spectrum. Paris, April 22-25, 2008.

\bibitem[2003]{Maraschi} Maraschi L. \& Tavecchio F.\ 2003, ApJ, 593, 667 

\bibitem[2001]{OM} Mason K.O., Breeveld A., Much R., et al., 2001, A\&A 365, L36

\bibitem[2008]{MASSARO} Massaro F., Tramacere A., Cavaliere A., et al., 2008, A\&A 478, 395

\bibitem[2007]{Mazin} Mazin D. \& Raue M., 2007, A\&A, 471, 439

\bibitem[2007]{Mazin2} Mazin D. \& Lindfors E., 2007, Proceedings 30$^{\rm th}$ ICRC, [\texttt{arXiv:0709.1694}]

\bibitem[2008]{CANGAROOIII} Sakamoto Y., Nishijima K., Mizukami T., et al., 2008, ApJ 676, 113

\bibitem[2001]{Struder01} Str\"uder L., Briel U., Dennerl K., et al., 2001, A\&A, 365, L18

\bibitem[2000]{Tagliaferri} Tagliaferri G., Ghisellini G., Giommi P., et al., 2000, A\&A 354, 431

\bibitem[2008]{FAB} Tavecchio F. \& Ghisellini G., 2008, MNRAS, accepted [\texttt{arXiv:0801.0593}]

\bibitem[2001]{Turner01} Turner M.J., Abbey A., Arnaud M., et al., 2001, A\&A, 365, L27

\bibitem[1982]{URRY1} Urry C.M. \& Mushotzky R.F., 1982, ApJ 253, 38

\bibitem[1995]{URRY3} Urry C.M. \& Padovani P., 1995, PASP 107, 803

\bibitem[1997]{URRY2} Urry C.M., Treves A., Maraschi L., et al., 1997, ApJ 486, 799
 
\bibitem[2008]{ZHANG} Zhang Y.H., 2008, ApJ, accepted [\texttt{arXiv:0804.3626}]

\end{thebibliography}
\end{document}